\documentclass{article}
\usepackage{spconf,amsmath,graphicx, multicol, hyperref}
\usepackage{lipsum}
\usepackage{epsfig}
\usepackage{epstopdf} 

\title{NIPS4Bplus: a richly annotated birdsong audio dataset}
\name{Veronica Morfi$^{\star}$ \qquad Yves Bas$^{\dagger,\ddagger}$ \qquad Hanna Pamu\l{}a $^{\dagger\dagger}$ \qquad Herv\'{e} Glotin $^{\ddagger\ddagger}$ \qquad Dan Stowell $^{\star}$ \thanks{We thank EADM MaDICS CNRS and SABIOD MI CNRS for supporting the NIPS4B challenge, Sylvain Vigant for providing recordings from Central France, and BIOTOPE for making the data public for the NIPS4B 2013 bird classification challenge.}}

\address{$^{\star}$Machine Listening Lab, Centre for Digital Music (C4DM), Queen Mary Univ. of London, UK\\$^{\dagger}$Centre d'Ecologie et des Sciences de la Conservation (CESCO), \\ Mus\'{e}um National d'Histoire Naturelle, CNRS, Sorbonne Univ., Paris, France\\$^{\ddagger}$Centre d'Ecologie Fonctionnelle et Evolutive (CEFE), CNRS, Univ. de Montpellier,\\ Univ. Paul-Val\'{e}ry Montpellier, France \\$^{\dagger\dagger}$AGH Univ. of Science and Technology, Department of Mechanics and Vibroacoustics, Krak\'{o}w, Poland\\$^{\ddagger\ddagger}$Univ. Toulon, Aix Marseille Univ., CNRS, LIS, DYNI team, SABIOD, Marseille, France}
\begin{document}
%
\maketitle
\begin{abstract}
Recent advances in birdsong detection and classification have approached a limit due to the lack of fully annotated recordings. In this paper, we present NIPS4Bplus, the first richly annotated birdsong audio dataset, that is comprised of recordings containing bird vocalisations along with their active species tags plus the temporal annotations acquired for them. Statistical information about the recordings, their species specific tags and their temporal annotations are presented along with example uses. NIPS4Bplus could be used in various ecoacoustic tasks, such as training models for bird population monitoring, species classification, birdsong vocalisation detection and classification. 
\end{abstract}
\section{Introduction}
\label{sec:intro}
The potential applications of automatic species detection and classification of birds from their sounds are many (e.g. ecological research, biodiversity monitoring, archival) \cite{Dawson:09, Lambert:14, Drake:16, Sovern:14, Marques:12}. In recent decades, there has been an increasing amount of ecological audio datasets that have tags assigned to them to indicate the presence or not of a specific bird species. Utilising these datasets and the provided tags, many authors have proposed methods for bird audio detection \cite{Adavanne:17a, Pellegrini:17} and bird species classification, e.g. in the context of LifeCLEF classification challenges \cite{Goeau:16, Goeau:17} and more \cite{Salamon:17, Knight:17}. However, these methods do not predict any information about the temporal location of each event or the number of its occurrences in a recording.  

Some research has been made into using audio tags in order to predict temporal annotations, labels that contain temporal information about the audio events. In \cite{Briggs:12, Ruiz:15}, the authors try to exploit audio tags in birdsong detection and bird species classification, in \cite{Fanioudakis:17}, the authors use deep networks to tag the temporal location of active bird vocalisations, while in \cite{Roger:18}, the authors propose a bioacoustic segmentation based on the Hierarchical Dirichlet Process (HDP-HMM) to infer song units in birdsong recordings. Furthermore, some methods for temporal predictions by using tags have been proposed for other types of general audio \cite{Schluter:16, Adavanne:17b, Kumar:16}. However, in all the above cases some kind of temporal annotations were used in order to evaluate the performance of the methods.

Annotating ecological data with temporal annotations to train sound event detectors and classifiers is a time consuming task involving a lot of manual labour and expert annotators. There is a high diversity of animal vocalisations, both in the types of the basic syllables and in the way they are combined \cite{ScottBrandes:08, Kroodsma:05}. Also, there is noise present in most habitats, and many bird communities contain multiple bird species that can potentially have overlapping vocalizations \cite{Luther:08, Luther:09a, Pacifici:08}. These factors make detailed annotations laborious to gather, while on the other hand acquiring audio tags takes much less time and effort, since the annotator has to only mark the active sound event classes in a recording and not their exact boundaries. This means that many ecological datasets lack temporal annotations of bird vocalisations even though they are vital to the training of automated methods that predict the temporal annotations which could potentially solve the issue of needing a human annotator. 

Recently, BirdVox-full-night \cite{Birdvox}, a dataset containing some temporal and frequency information about flight calls of nocturnally migrating birds, was released. However, BirdVox-full-night only focuses on avian flight calls, a specific type of bird calls, that usually have a very short duration in time. The temporal annotations provided for them don't include any onset, offset or information about the duration of the calls, they simply contain a single time marker at which the flight call is active. Additionally, there is no distinction between the different bird species, hence no specific species annotations are provided, but only the presence of flight calls through the duration of a recording is denoted. Hence, the dataset can provide data to train models for flight call detection but is not efficient for models performing both event detection and classification for a variety of bird vocalisations. 

In this paper, we introduce NIPS4Bplus, the first ecological audio dataset that contains bird species tags and temporal annotations \cite{Morfi:18data}, and can be used for training supervised automated methods that perform bird vocalisation detection and classification and can also be used for evaluating methods that use only audio tags or no annotations for training. The rest of the paper is structured as follows: Section \ref{sec:dataset} describes the process of collecting and selecting the recordings comprising the dataset, Section \ref{sec:anno} presents our approach of acquiring the tags and temporal annotations and provides statistical information about the labels and recordings comprising the dataset followed by example uses of NIPS4Bplus, with the conclusion in Section \ref{sec:conclusion}.

\section{Audio Data Collection}
\label{sec:dataset}

\begin{center}
\begin{figure}[!htb]
\centering
  \includegraphics[width=7cm]{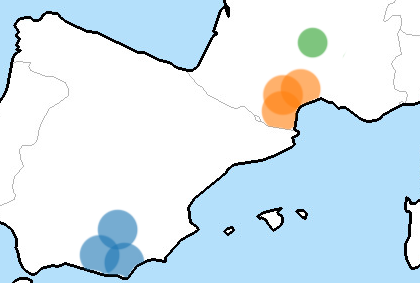}
  \caption{Regions where the dataset recordings were collected from. Green indicates Central France region Haute-Loire. Orange indicates Southern France regions Pyr\'{e}n\'{e}es-Orientales, Aude and H\'{e}rault. Blue indicates Southern Spain regions Granada, Ja\'{e}n and Almeria.}
  \label{fig:1}
\end{figure}
\end{center}

The recordings that comprise the Neural Information Processing Scaled for Bioacoustics (NIPS4B) 2013 training and testing dataset were collected by recorders placed in 39 different locations, which can be summarised by 7 regions in France and Spain, as depicted in Fig. \ref{fig:1}. 20\% of the recordings were collected from the Haute-Loire region in Central France, 65\% of them were collected from the Pyr\'{e}n\'{e}es-Orientales, Aude and H\'{e}rault regions in south-central France along the Mediterranean cost and the remaining 15\% of the recordings originated from the Granada, Ja\'{e}n and Almeria regions in eastern Andalusia, Spain. The Haute-Loire area is a more hilly and cold region, while the rest of the regions are mostly along the Mediterranean coast and have a more Mediterranean climate. 

The recorders used to acquire the recordings were the SM2BAT using SMX-US microphones. They were originally put in the field for bat echolocation call sampling, but they were also set to record for 3 hours single channel at 44.1 kHz sampling rate starting 30 minutes after sunrise, right after bat sampling. The recorders were set to a 6 dB Signal-to-Noise Ratio (SNR) trigger with a window of 2 seconds, and acquired recordings only when the trigger was activated.

Approximately 30 hours of field recordings were collected. Any recording longer than 5 seconds was split into multiple 5 second files. SonoChiro, a chirp detection tool used for bat vocalisation detection, was used on each file to identify recordings with bird vocalisations.\footnote{\url{http://www.leclub-biotope.com/fr/72-sonochiro}} A stratified random sampling was then applied to all acquired recordings, based on locations and clustering of features, to maximise the diversity in the labelled dataset, resulting in nearly 5000 files being chosen. Following the first stage of selection, manual annotations were produced for the classes active in these 5000 files and any recordings that contained unidentified species' vocalisations were discarded. Furthermore, the training set and testing set recordings were allocated so that the same species were active in both. Finally, for training purposes, only species that could be covered by at least 7 recordings in the training set were included in the final dataset, the rest were considered rare species' occurrences that would make it hard to train any classifier, hence were discarded. The final training and testing set consist of 687 files of total duration of less than an hour, and 1000 files of total duration of nearly two hours, respectively.
 
\vspace{-1.2cm}

\section{Annotations}
\label{sec:anno}

\subsection{Tags}
\label{ssec:weak}
\begin{figure*}[!htb]
  \includegraphics[width=18cm]{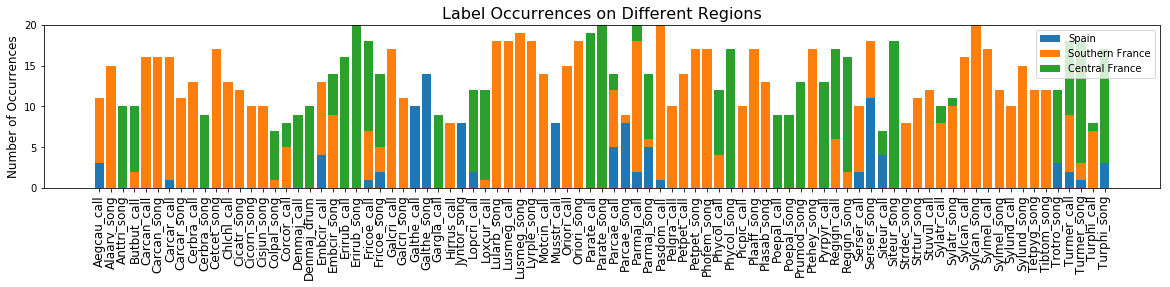}
  \caption{Number of occurrences of each sound type in recordings collected from Spain, Southern France and Central France.}
  \label{fig:2}
\end{figure*}

The labels for the species active in each recording of the training set were initially created for the NIPS4B 2013 bird song classification challenge \cite{NIPS4B2013}. There is a total of 61 different bird species active in the dataset. For some species we discriminate the song from the call and from the drum. We also include some species living with these birds: 7 insects and an amphibian. This tagging process resulted in 87 classes. A detailed list of the class names and their corresponding species English and scientific names can be found in \cite{Morfi:18data}. These tags only provide information about the species active in a recording and do not include any temporal information. In addition to the recordings containing bird vocalisations, some training files only contain background noise acquired from the same regions and have no bird song in them, these files can be used to tune a model during training. Fig. \ref{fig:2} depicts the number of occurrences per class for recordings collected in each of the 3 different general regions of Spain, South France and Central France. Each tag is represented by at least 7 up to a maximum of 20 recordings.

\begin{figure}[!htb]
  \includegraphics[width=8cm]{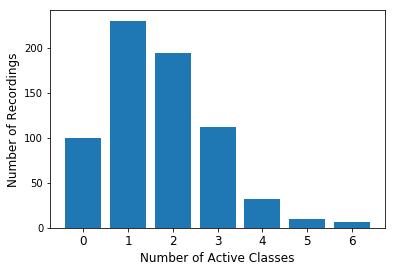}
  \caption{Distribution of number of active classes in dataset recordings.}
  \label{fig:3}
\end{figure}

Each recording that contains bird vocalisations includes 1 to 6 individual labels. These files may contain different vocalisations from the same species and also may contain a variety of other species that vocalise along with this species. Fig. \ref{fig:3} depicts the distribution of the number of active classes in the dataset. 

\vspace{-0.5cm}

\subsection{Temporal Annotations}
\label{ssec:trans}
\begin{figure}[!htb]
  \includegraphics[width=8.5cm]{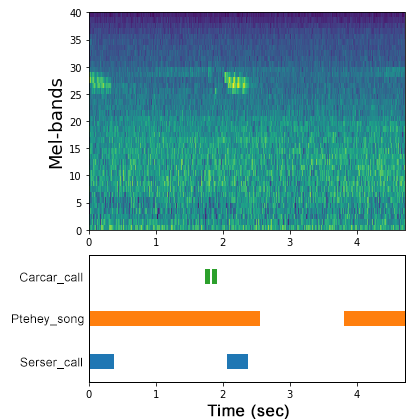}
  \caption{Mel-band spectrogram of a recording in NIPS4Bplus and the visual representation of the corresponding temporal annotations as noted in Table \ref{tab:1}.}
  \label{fig:5}
\end{figure}

Temporal annotations for each recording in the training set of the NIPS4B dataset were produced manually using Sonic Visualiser.\footnote{\url{https://www.sonicvisualiser.org/}} The temporal annotations were made by a single annotator, Hanna Pamu\l{}a, and can be found in \cite{Morfi:18data}. Table \ref{tab:1} presents the temporal annotation format as is provided in NIPS4Bplus and Fig. \ref{fig:5} depicts the visual representation of the temporal annotations.

In concern to the temporal annotations for the dataset, we should mention the following:
\begin{itemize}
\vspace{-0.2cm}
\item The original tags were used for guidance, however some files were judged to have a different set of species than the ones given in the original metadata.
\vspace{-0.2cm}
\item In a few rare occurrences, despite the tags suggesting a bird species active in a recording, the annotator was not able to detect any bird vocalisation.
\vspace{-0.2cm}
\item An extra `Unknown' tag was added to the dataset for vocalisations that could not be classified to a class.
\vspace{-0.2cm}
\item An extra `Human' tag was added to a few recordings that have very obvious human sounds, such as speech, present in them.
\vspace{-0.2cm}
\item Out of the 687 recordings of the training set 100 recordings contain only background noise, hence no temporal annotations were needed for them.
\vspace{-0.2cm}
\item Of the remaining 587 recordings that contain vocalisations, 6 could not be unambiguously labelled due to hard to identify vocalisations, thus no temporal annotation files were produced for them. 
\vspace{-0.2cm}
\item An annotation file for any recording containing multiple insects does not differentiate between the insect species and the `Unknown' label was given to all insect species present.
\vspace{-0.2cm}
\item In the rare case where no birds were active along with the insects no annotation file was provided. Hence, 7 recordings containing only insects were left unlabelled.
\vspace{-0.6cm}
\item In total, 13 recordings have no temporal annotation files. These can be used when training a model that does not use temporal annotations.
\vspace{-0.2cm}
\item On some occasions, the different syllables of a song were separated in time into different events while in other occasions they were summarised into a larger event, according to the judgement of the expert annotator. This variety could help train an unbiased model regarding separating events or grouping them together as one continuous time event.
\end{itemize}

\begin{table}[!htb]
\caption{NIPS4Bplus temporal annotations of the recording depicted in Fig. \ref{fig:5}}
\begin{center}
\begin{tabular}{ |c c c| } 
\hline
\textbf{Starting Time (sec)} & \textbf{Duration (sec)} & \textbf{Tag} \\
\hline
0.00 & 0.37 & Serser\_call \\
0.00 & 2.62 & Ptehey\_song\\
1.77 & 0.06 & Carcar\_call\\
1.86 & 0.07 & Carcar\_call\\
2.02 & 0.41 & Serser\_call\\
3.87 & 1.09 & Ptehey\_song\\
\hline
\end{tabular}
\end{center}
\label{tab:1}
\end{table}

\begin{figure}[t]
  \includegraphics[width=8cm]{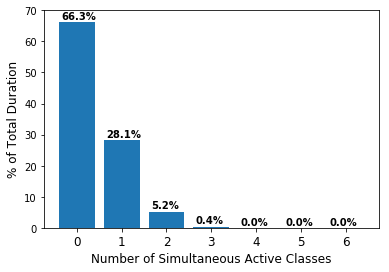}
  \caption{Distribution of simultaneous number of active classes on the total duration of the recordings.}
  \label{fig:4}
\end{figure}

As mentioned above, each recording may contain multiple species vocalising at the same time. This can often occur in wildlife recordings and is important to be taken into account when training a model. Fig. \ref{fig:4} presents the fraction of the total duration containing overlapping vocalisations as well as the number of simultaneously occurring classes. 

A few examples of the NIPS4Bplus dataset and temporal annotations being used can be found in \cite{Morfi:18b} and \cite{Morfi:18a}. First, in \cite{Morfi:18b}, we use NIPS4Bplus to carry out the training and evaluation of a newly proposed multi-instance learning (MIL) loss function for audio event detection. And in \cite{Morfi:18a}, we combine the proposed method of \cite{Morfi:18b} and a network trained on the NIPS4Bplus tags that performs audio tagging in a multi-task learning (MTL) setting. Additional applications using NIPS4Bplus could include training models for bird species audio event detection and classification, evaluating how generalisable of method trained on a different set of data is, and many more.

\section{Conclusion}
\label{sec:conclusion}
In this paper, we present NIPS4Bplus, the first richly annotated birdsong audio dataset. NIPS4Bplus is comprised of the NIPS4B dataset and tags used for the 2013 bird song classification challenge plus the newly acquired temporal annotations. We provide statistical information about the recordings, their species specific tags and their temporal annotations. 

%


\bibliographystyle{IEEEbib}
\bibliography{strings,refs,stage2_bibliography}

\end{document}